\documentclass[12pt]{article}
\usepackage{amsmath}
\usepackage{amssymb}
\usepackage{amsfonts}
\usepackage{amsbsy}
\usepackage{cite}
\usepackage{graphicx}
\usepackage[font=small,labelfont=bf,labelsep=period]{caption}

\usepackage{latexsym} 
\usepackage{slashed} 

\begin{document}

\addcontentsline{toc}{subsection}{{Title of the article}\\
{\it B.B. Author-Speaker}}

\setcounter{section}{0}
\setcounter{subsection}{0}
\setcounter{equation}{0}
\setcounter{figure}{0}
\setcounter{footnote}{0}
\setcounter{table}{0}

\begin{center}
\textbf{RECURSIVE FRAGMENTATION MODEL WITH QUARK SPIN. APPLICATION TO QUARK POLARIMETRY%
\footnote{
presented at XIII$^{\rm th}$ WORKSHOP ON HIGH ENERGY SPIN PHYSICS, 
\emph{DSPIN-2009}, Dubna, Russia, Sept. 1-5, 2009 ; http://theor.jinr.ru/~spin/2009/ 
}}

\vspace{5mm}
X. Artru 
\vspace{5mm}

\begin{small}
\emph{Institut de Physique Nucl\'eaire de Lyon, Universit\'e de Lyon,\\ Universit\'e Lyon 1 and CNRS/IN2P3, F-69622 Villeurbanne, France} \\
\emph{E-mail: x.artru@ipnl.in2p3.fr}
\end{small}
\end{center}

\vspace{0.0mm} 

\begin{abstract}
An elementary recursive model accounting for the quark spin in the fragmentation of a quark into mesons is presented. The quark spin degree of freedom is represented by a two-components spinor. Spin one meson can be included. The model produces Collins effect and jet handedness. The influence of the initial quark polarisation decays exponentially with the rank of the meson, at different rates for longitudinal and transverse polarisations. 
\end{abstract}

\vspace{3mm} 

%
\def\vold#1{\boldsymbol{#1}}
\def\RE{\mathop{\Re{\rm e}}\nolimits}
\def\IM{\mathop{\Im{\rm m}}\nolimits}
\def\Tr{\mathop{\rm Tr}\nolimits}
\def\un{\mathbf{1}}
%
%
%
\def\vsigma{\vold\sigma}
\def\ver{\Gamma}
\def\pro{\Delta}
\def\d{\mathrm{d}} 
\def\sv{{\vold{s}}}
\def\av{{\vold{a}}}
\def\pv{{\vold{p}}}
\def\qv{{\vold{q}}}
\def\rv{{\vold{r}}}
\def\kv{{\vold{k}}}
\def\Av{{\vold{A}}}
\def\Gv{{\vold{G}}}
\def\Pv{{\vold{P}}}
\def\Sv{{\vold{S}}}
\def\Jv{{\vold{J}}}
\def\Ev{{\vold E}}
\def\Bv{{\vold B}}
\def\Rv{{\vold R}}
\def\Pv{{\vold P}}
\def\Kv{{\vold K}}
\def\Qv{{\vold Q}}
\def\Vv{{\vold V}}
\def\ku{\hat{\vold{k}}} 
\def\pu{\hat{\vold{p}}}
\def\zu{\rm\bf\hat z}
\def\xu{\rm\bf\hat x}
\def\yu{\rm\bf\hat y}
\def\T{{\rm T}}
\def\Mcal{{\mathcal{M}}}
\def\Ma{{\rm\bf M}}
\def\Fcal{{\mathcal{F}}}
\def\Hcal{{\cal H}}
\def\Ocal{{\mathcal{O}}}
\def\Rcal{{\mathcal{R}}}
\def\Rt{{\mathcal{\widetilde{R}}}}
\def\Rh{{\mathcal{\widehat{R}}}}
\def\be{\begin{equation}}
\def\ee{\end{equation}}
\def\ni{\noindent}

\def\qt{{\vold{t}}}
\def\qtil{\tilde{\vold{t}}}
\def\ql{l}
\def\qlil{{\tilde{l}}}
\def\qq{\bar q_{-1}} 
\def\barqq{q_{-1}}
\def\q{q_0}

\section{Introduction}

Present Monte-Carlo event generators of quark and gluon jets do not include the parton spin degree of freedom, therefore do not generate the Collins \cite{Collins} and jet handedness \cite{handedness} effects. These are azimuthal asymmetries bearing on one, two or three hadrons, which can serve as \emph{quark polarimeters}. However the asymmetries may strongly depend, in magnitude and sign, on the quark and hadron flavors and on the transverse momenta $\pv_T$ and scaled longitudinal momenta $z$ of these hadrons. Therefore a good knowledge of this dependence is needed for parton polarimetry. Due to the large number of kinematical variables, a hadronisation model which takes spin into account is urgently needed as a guide. 

The semi-classical Lund $^3P_0$ mechanism \cite{Lund}, grafted on the string model, can generate a Collins effect \cite{ACY}, but not jet-handedness. Here we propose a fully quantum model of spinning quark fragmentation, based on the multiperipheral model. It reproduces the results of the $^3P_0$ mechanism and also contains the jet-handedness effect.

\section{Some recalls about quark fragmentation}

Figure 1 describes the creation of a quark "$\q$" and an antiquark "$\qq$" in $e^+e^-$ annihilation or $W^\pm$ decay, followed by the hadronisation,
\be\label{hadronis} 
\q+\qq \to h_1+h_2...+h_N
~.\ee
Looking from rigth to left, one sees it as the recursive process (see \cite{recursive} and ref. 4 of \cite{universal}), 
\be\label{cascade} 
\begin{array}{l}
\q\equiv q_0 \to h_1+q_1\\ 
\quad\  \ \  q_{1} \to\ h_2+q_2 \\
 \quad \qquad \cdots  \\
\ q_{N-1} \to\ h_N+q_N
\end{array} 
\qquad
\begin{array}{r}
\hbox{4-momenta :}\qquad
k_{0}=p_{1}+k_{1}~, \\
k_{1}=p_{2}+k_{2}~, \\
 \cdots \qquad\quad \\
 k_{N-1}=p_{N}+k_{N}~.
\end{array} 
\ee
%

\begin{figure} 
  \centering 
 \vspace*{-10mm} 
  \includegraphics [width=100mm]
 {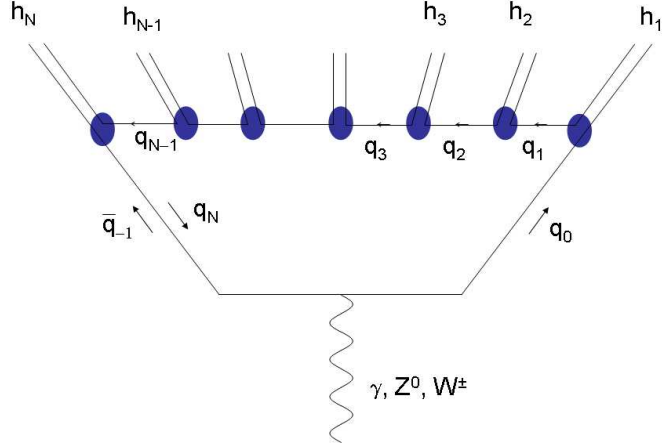}
  \caption{\footnotesize Electroweak boson $\to q\bar q\to$ mesons.}
\end{figure}

\ni $q_N\equiv\barqq$ is a "quark propagating backward in time" and $k_N\equiv-k(\qq)$. 

\medskip\ni{\it Kinematical notations~:} 
\par\ni
$\kv_0=\kv(\q)$ and $\kv(\qq)$ are in the $+\zu$ and $-\zu$ directions respectively. For a quark, $\qt_n\equiv\kv_{nT}$. For a 4-vector, ~$a^\pm=a^0\pm a^z$ and $\av_T=(a^x,a^y)$. 
We denote by a tilde the \emph{dual transverse vector}
$\tilde{\av}_T\equiv\zu\times\av_T=(-a^y,a^x)$.

\medskip\ni In Monte-Carlo simulations, the $k_n$ are generated according to the \emph{splitting distribution} 
\be\nonumber 
dW(\ q_{n-1} \to\ h_n+q_n)=f_{n}(\zeta_n,\qt_{n-1}^2,\qt_n^2,\pv_{nT}^2,)\, d\zeta_n\,d^2\qt_n~,\qquad \zeta_n\equiv p^+_n/k^+_{n-1}
~.\ee
In particular the \emph{symmetric Lund} splitting function \cite{Lund},
\be  \label{L-Sff} 
f_{n} \propto
\zeta_n^{a_{n-1}-a_n-1}\,(1-\zeta^{a_n})\,\exp\left[-b\,(m_n^2+\pv_{nT}^2)/\zeta_n\right]
~,\ee
inspired by the string model, fulfills the requirement of \emph{forward-backward equivalence}. 

On can also consider \cite{universal} the upper part of Fig.1 as a \textbf{multiperipheral} \cite{multiperiph} diagram with the Feynman amplitude
\begin{eqnarray}\label{multiperampli} 
	\Mcal_{\q+\qq \to h_1...+h_N} = \bar v_(k_{-1},\Sv_{-1}) \
\ver_{q_N,h_N,q_{N-1}}(k_N,k_{N-1}) \ \pro_{q_{N-1}}(k_{N-1}) \ \cdots  \cr 
	\cdots\	\pro_{q_{2}}(k_2) \	\ver_{q_{2},h_2,q_{1}}(k_2,k_{1}) \ 
	\pro_{q_{1}}(k_1) \ \ver_{q_1,h_1,q_{0}}(k_1,k_{0}) \ u(k_0,\Sv_0)
~.\end{eqnarray}
$\Sv_0$ and $\Sv_{-1}$ are the polarisation vectors of the intial quark and antiquark. $\Sv^2=1$,  $S_z=$ helicity, $\Sv_T=$  transversity. $\ver$ and $\pro$ are vertex functions and propagators which depend on the quark momenta and flavors. Note that Fig.1 is a loop diagram~: $k_0$ is an integration variable, therefore the "jet axis" is not really defined. Furthermore, in $Z_0$ or $\gamma^*$ decay, the spins $\q$ and $\qq$ are entangled so that one cannot define $\Sv_0$ and $\Sv_{-1}$ separately. 

\paragraph{Collins and jet-handedness effects.}

Let us first assume that the \emph{jet axis} (quark direction) is well determined~:

\medskip\ni
- the \emph{Collins effect} \cite{Collins}, in $\vec{q}\to h+X $, is an asymmetry in $\sin[\varphi(\Sv)-\varphi(h)]$ for a transversely polarized quark. The fragmentation function reads
\be \label{CE}  
F(z,\pv_T\,;\Sv_T)=F_0(z,\pv_T^2)\, 
\left(1+A_T\,\Sv_T.\tilde{\pv}_T/|\pv_T|\right) 
\qquad (\tilde{\pv}_T\equiv\zu\times\pv_T)
~.\ee
$A_T=A_T(z,\pv_T^2)\in[-1,+1]$ is the Collins analysing power. 

\medskip\ni 
- \emph{jet handedness} \cite{handedness}, in $\vec{q}\to h+h'+X $, is an asymmetry in $\sin[\varphi(h)-\varphi(h')]$ proportional to the quark helicity. The 2-particle longitudinaly polarised fragmentation function is
\be 
F(z,\pv_{T},z',\pv'_{T}\,;S_z) =
F_0(z,\pv_{T}^2,z',{\pv'_{T}}^2,\pv_{T}\cdot\pv'_{T})\,
\left(1+A_L\,S_z\,\frac{\tilde{\pv}_{T}.\pv'_{T}}{|\tilde{\pv}_{T}\cdot\pv'_{T}|}\right) 
\,. \label{JH}  
\ee
$A_L=A_L(z,\pv_{T}^2,z',{\pv'_{T}}^2,\pv_{T}.\pv'_{T})\in[-1,+1]$ is the handedness analysing power. ~$\tilde{\pv}_{1T}\cdot\pv'_{T}$ is the same as $\zu\cdot(\pv_{T}\times\pv'_{T})$.

\medskip
If the jet axis is not well determined, an additional fast hadron, $h'$ or $h''$ is needed. The $z$ axis is taken along $\Pv=(\pv+\pv')$ (Collins) or $\Pv=(\pv+\pv'+\pv'')$ (handedness). In this way, we define the \emph{2-particle relative Collins effect} (also called \emph{interference fragmentation}) and the \emph{three-particle} jet handedness, which corresponds to the original definition of \cite{handedness}. 

\begin{figure} 
  \centering 
\vspace*{-30mm} 
\includegraphics [width=140mm]
 {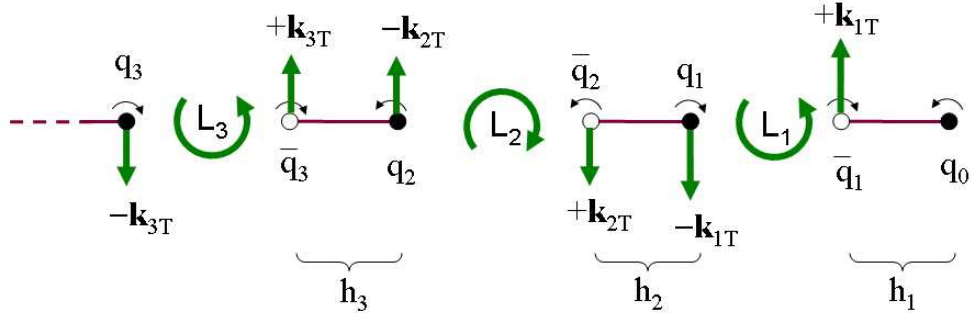}
\vspace*{-20mm}
 \caption{\footnotesize String decaying into pseudoscalar mesons.}
\end{figure}

\paragraph{The Lund $^3P_0$ mechanism\cite{Lund}.}

Figure 2 depicts the decay of the initial massive string accompagnied with the creation of a $q\bar q$ pairs. Forgetting transverse oscillations of the initial string, the transverse hadron momenta come from the internal orbital motions of the pairs. After a tunnel effect the $q$ and $\bar q$ of a pair become on-shell and their relative position $\rv\equiv\rv(q)-\rv(\bar q)$ is along $-\zu$. The pair is assumed to be in the $^3P_0$ state, which has the vacuum quantum number. The relative momentum $\kv\equiv\kv(q)=-\kv(\bar q)$ and the orbital angular momentum $\vold{L}=\rv\times\kv$ are such that  
$\zu\cdot[\kv_T\times\vold{L}]<0$. In the $^3P_0$ state $\langle\sv_q\rangle = \langle\sv_{\bar q}\rangle = -\langle\vold{L}/2\rangle$. As a result, the transverse spins of $q$ and $\bar q$ are correlated to their transverse momenta~:
\be \label{Lu-meca} 
\langle\, \zu\cdot[\kv_T(q)\times\sv_q]\,\rangle >0 
\,,\qquad
\langle\, \zu\cdot[\kv_T(\bar q)\times\sv_{\bar q}]\,\rangle <0 \,.
\ee
The correlation can be transmitted to a baryon. Then $\langle\, \zu\cdot[\pv_T\times\sv_B]\,\rangle$ has the sign of $\langle\sv_{q}.\sv_{B}\rangle$. This can explain transverse spin asymmetries in hyperon production \cite{Lund}. 

\medskip
\ni \emph{Application to the Collins effect} \cite{ACY}.~~
In Fig. 2, $q_0$ is polarised along the direction $+\yu$ toward the reader and $h_1$ is a pseudoscalar meson, for which $\langle\sv(q_0)\rangle=-\langle\sv(\bar q_1)\rangle$. Then $q_1$ and $\bar q_1$ are polarised along $-\yu$ and, according to (\ref{Lu-meca}), $\kv_T(\bar q_1)=\pv_T(h_1)$ is in the  $+\xu$ direction. This provides a model for the Collins effect. Fig. 2 also indicates that, for a sequence of pseudoscalar mesons, the Collins asymmetries are of alternate sides. Besides, $q_{n-1}$ and $\bar q_{n}$ go on the same side, which enhances the asymmetry. It may explain why $\pi^-$ from $u$-quarks have a strong Collins analysing power. Note that this effect also enhances $\langle\pv_T^2\rangle$, independently of the $q_0$ polarisation.  
 
\section{A simplified multiperipheral quark model}

In Eq.(\ref{multiperampli}), let us replace Dirac spinors by Pauli spinors. A minimal model, restricted to the direct emission of pseudoscalar mesons, is built with the following prescriptions~:

\smallskip\ni
1) replace $u(k_0,\Sv_0)$ and $\bar v(k_{-1},\Sv_{-1})\equiv-\bar u(k_{\qq},-\Sv_{\qq})\,\gamma_5$  by the Pauli spinors $\chi(\Sv_0)$ and 
\par\ni~~~ $-\chi^\dagger(-\Sv_{-1})\,\sigma_z$~, 
\par\smallskip\ni 
2) assume no momentum dependence of $\ver$~,
\par\smallskip\ni 
3) replace $\gamma_5$ by $\sigma_z$~,
\par\smallskip\ni 
4) replace the usual pole $(k^2-m_q^2)^{-1}$ of  $\pro_{q}(k) $ by the $(k_L,\,k_T)$ separable form
	\be
	D_q(k)=g_q(k^+k^-)\,\exp(-B\qt^2/2)~,
	\ee
\par\ni 5) replace the usual numerator ~$m_q+\gamma.k$~ by
~$\mu_q(k^+k^-,\qt^2)+i\vsigma.\qtil$~.
\par\smallskip\ni
These prescriptions respect the invariance under the following transformations~:
	\par\smallskip\ni ~- rotation about the $z$-axis,
	\par\smallskip\ni ~- Lorentz transformations along the $z$-axis (longitudinal boost), 
	\par\smallskip\ni ~- mirror reflection about any plane containing the $z$-axis (parity), 
	\par\smallskip\ni ~- forward-backward equivalence.
\par\smallskip\ni	
The jet axis being fixed, full Lorentz invariance is not required, whence the separate dependences of $D_q$ and $\mu_q$ in $k^+k^-$ and $\qt^2$. In item 5), ~$\mu+i\vsigma.\qtil$~ is reminiscent of the meson-baryon scattering amplitude $f(s,t)+ig(s,t)\,\vsigma.(\pv\times\pv')$. Single-spin effects are obtained for $\IM\mu\ne0$. The choice of putting the spin dependence in the propagators rather than in the vertices is inspired by the $^3P_0$ mechanism~: in both models the polarisation germinates in the quark line between two hadrons. 

For a fast investigation of the model, we make the further approximations :

\smallskip
\ni - \emph{Neglect the influence of the antiquark flavor and polarisation in the quark fragmentation region.} This is allowed at large invariant $\q+\qq$ mass.

\smallskip
\ni - \emph{Discard the interference diagrams}. 
For a given final state, the \emph{rank ordering} of hadrons in the multiperipheral diagram is not unique and differently ordered diagrams can interfere. This interference (and the resulting Bose-Einstein correlations) will be neglected.

\smallskip
\ni - \emph{Disentangle $k^\pm$ and $\kv_T$.} We will assume that $\mu_q(k^+k^-,\qt^2)$ is constant or a function of $\qt^2$ only. Thus we have no more "dynamical" correlation between longitudinal and transverse momenta. However there remains a "kinematical" correlation coming from the mass shell constraint 
\be  \label{shell} 
(k_{n-1}-k_{n})^2\equiv(k^+_{n-1}-k^+_{n})(k^-_{n-1}-k^-_{n})-(\qt_{n-1}-\qt_{n})^2
=m^2(h_n)
~.\ee
In the following we will ignore the $(\qt_{n-1}-\qt_{n})^2$ term. This approximation is drastic for pion emission because  $\langle \qt^2\rangle>m_\pi^2$. We only use it here for a qualitative investigation of the spin effects allowed by the multiperipheral model. Thanks to it, the $\qt$'s become fully decoupled from the $k^\pm_n$ and kinematically decorrelated between themselves. They remain correlated only via the quark spin.

\paragraph{$\pv_T$-distibutions in the quark fragmentation region.}

With the above approximations we can treat the process (\ref{cascade}), at least in $\pv_T$-space, like a cascade decay of unstable particles, which has no constraint coming from the future. The joint $\pv_T$ distibution of the $n$ first mesons is proportional to
\be\label{factoris} 
I(\pv_{1T},\pv_{2T},...\pv_{nT}) =\exp(-B\,\qt_1^2-B\,\qt_2^2\,...-B\,\qt_n^2)\,
\Tr\left\{\Ma_{12...n}\,\frac{\un+\Sv_0.\vsigma}{2}\,{\Ma}_{12...n}^\dagger\right\}
\,,
\ee
with
\be\label{Ma} 
\Ma_{12...n}= \Ma_n \cdots\Ma_2\,\Ma_1
\,,\qquad
\Ma_r=(\mu_r+i\vsigma.\qtil_r)\,\sigma_z 
~.\ee

\subsection{Applications to azimuthal asymmetries}

In this section we will calculate azimuthal asymmetries for particles of definite ranks. For comparison with experiments, one should mix the contributions of different rank assignments. For simplicity we take a unique and constant $\mu$ for all quark flavors. 

\paragraph{First-rank Collins effect.}

Applying (\ref{factoris}-\ref{Ma}) for $n=1$ gives
\be \label{rank1} 
I(\pv_{1T}) = \exp(-B\qt_1^2)\,\left(\,|\mu|^2+\qt_1^2-2\IM(\mu)\,\qtil_1.\Sv\,\right)
\,,\ee
with $\qt_1=-\pv_{1T}$. For complex $\mu$ one 
has a Collins asymmetry (cf Eq.\ref{CE}) with
\be \label{A-Col} 
A_T= 2\frac{\IM(\mu)\,|\pv_{1,T}|}{|\mu|^2+\pv_{1,T}^2} \ \in[-1,+1]
\,. \ee
If $\IM(\mu)>0$ it has the same sign as predicted by the $^3P_0$ mechanism.

\paragraph{Joint $\pv_T$ spectrum of $h_1$ and $h_2$.} 
Applying (\ref{factoris}-\ref{Ma}) for $n=2$ one obtains 
\begin{eqnarray} \label{rank2} 
I(\pv_{1T},\pv_{2T}) &=& \exp(-B\qt_1^2-B\qt_2^2)\, 
\{\,(|\mu|^2+\qt_1^2)\,(|\mu|^2+\qt_2^2)-4\qt_1.\qt_2\,\IM^2(\mu) 
\cr &+&
2\IM(\mu)\,\Sv.\qtil_1 \,(2\,\qt_1.\qt_2-|\mu|^2-\qt_2^2)
\cr &+& 2\IM(\mu)\,\Sv.\qtil_2\,(|\mu|^2-\qt_1^2)
\cr &-& 2\IM(\mu^2)\,\Sv.(\qt_1\times\qt_2)
\,\}\,, 
\end{eqnarray}
with $\qt_1=-\pv_{1T}$, ~$\qt_2=-(\pv_{1T}+\pv_{2T})$ and 
$\Sv\cdot(\qt_1\times\qt_2)=S_z\ \tilde{\pv}_{1T}\cdot\pv_{2T}$.

The last line contains jet handedness (cf Eq.\ref{JH}), of analysing power 
\be \label{A_L} 
A_L= \frac{-2\IM(\mu^2)\,|\pv_{1T}\times\pv_{2T}|
}{
(|\mu|^2+\qt_1^2)\,(|\mu|^2+\qt_2^2)-4\qt_1.\qt_2\,\IM^2(\mu)
}\ \in[-1,+1]
\,. \ee
The second line contains the Collins asymmetry of $h_1$. Both $2^{\rm nd}$ and $3^{\rm rd}$ lines contribute to the $h_2$ one after integration over $\qt_1$, and to the \emph{relative 2-particle Collins asymmetry}, which bears on 
\be \label{ptrel} 
\rv_{12}= \frac{z_2\pv_{1T}-z_1\pv_{2T}}{ z_1+z_2} = \frac{z_1}{ z_1+z_2}\,\qt_2-\qt_1
~.\ee
Note that Collins and jet-handedness asymmetries are not maximum for the same value of $\arg(\mu)$. This is related to the \emph{positivity} \cite{inequal} constraint
\be \label{positiv} 
A_L^2(\pv_{1T},\pv_{2T}) +A_T^2(\pv_{1T},\pv_{2T})\le1
~.\ee

\subsection{Evolution of the polarisation of the cascading quark }
Let us first assume that $\qt_1$, $\qt_2$, ... $\qt_n$ are fixed and consider the spin density matrix $\rho_n=(\un+\Sv_n.\vsigma)/2$ of $q_n$ at the $(n+1)^{\rm th}$ vertex~: 
\be  \label{rhon} 
\rho_n = R_n/\Tr\{R_n\}\,,\qquad
R_n =  \Ma_{12...n}\ \frac{\un+\Sv_0.\vsigma}{2}\, {\Ma}_{12...n}^\dagger
~.\ee
If $\rho_0$ is a pure state (${\rm det}\, \rho_0=0)$, then $\rho_n$ is also a pure state~; no information is lost.   

Let us now integrate over $\qt_1$, $\qt_2$, ... $\qt_n$ (equivalently over $\pv_{1T}$, ...$\pv_{nT}$). It leads to a loss of information. The spin density matrix of $q_n$ becomes 
\be  \label{rhonbar} 
\bar\rho_n = \bar R_n/\Tr\{\bar R_n\}\,,\qquad
\bar R_n =  
\int d^2\qt_1...\int d^2\qt_{n}\ \Ma_{12...n}\ \frac{\un+\Sv_0.\vsigma}{2}\,{\Ma}_{12...n}^\dagger
~.\ee
$R_n$ and $\bar R_n$ obey the recursion relations
\be  \label{evol-rho} 
R_n = \Ma_{n}\,R_{n-1}\,{\Ma}_{n}^\dagger
~,\qquad
\bar R_n = \int d^2\qt_n \, \Ma_{n}\,\bar R_{n-1}\,{\Ma}_{n}^\dagger
~.\ee
At fixed $\qt$'s, the left equation gives (setting $\mu=\mu'+i\mu''$)~:
\be  \label{evol-S} 
\Sv_n = \frac{1}{C} \bigg\{ 2\mu'' \,\qtil_n +
\Rcal[\zu,\varphi_n] 
\begin{pmatrix}
\qt^2-|\mu|^2 & 0 &- 2 |\qt|\,\mu' \cr
0 & -|\mu|^2- \qt^2  & 0\cr
  2 |\qt|\,\mu'& 0 & |\mu|^2 - \qt^2 
\end{pmatrix} 
\Rcal[\zu,-\varphi_n] \,\Sv_{n-1} \bigg\}
~,\ee
with $C=\Tr\{R_n\}=|\mu|^2+\qt_n^2-2\mu''\,\qtil_{n}\cdot\Sv_{n-1}$. The rotation $\Rcal[\zu,\varphi_n]$ about $\zu$ brings $\xu$ along $\qt_n$. 
Iterating (\ref{rank1}), where we replace $\{\Sv,\,\qt_{1}\}$ by $\{\Sv_{n-1},\,\qt_{n}\}$, and (\ref{evol-S}), we generate the successive transverse momenta with the Monte-Carlo method. From (\ref{evol-S}) we learn~:
\par\medskip
\ni ~- if $\IM\mu\ne0$, the inhomogeneous term in $\mu''\,\qtil_n$ is a source (or sink) of transverse 
\par\ni~~~polarisation~: one can have $\Sv_{nT}\ne0$ even with $\Sv_{n-1}=0$.
\par\ni ~- helicity is partly converted into transversity along $\qt_n$ and vice-versa. 
\par\medskip
\ni The last fact explains the mechanism of jet handedness in this model~: first, the helicity $S_{z0}$ is partly converted into $\Sv_{1T}$ parallel to $\pv_{1T}$, then $\Sv_{1T}$ produces a Collins asymmetry for $h_2$ in the plane perpendicular to $\pv_{1T}$.

\medskip\ni
Let us now consider the $\qt$-integrated density matrix. The right equation in (\ref{evol-rho}) gives
\be \label{attenu} 
S_{n,z}= D_{LL} \, S_{n-1,z}~, \qquad \Sv_{n,T}= D_{TT} \, \Sv_{n-1,T}
~; \qquad D_{LL},\,D_{TT} \in[-1,+1]
\,,\ee
\be \label{attenu-SzST} 
\begin{pmatrix}
	D_{LL} \\
	D_{TT}
\end{pmatrix}
=\int d^2\qt\, \exp(-B\qt^2)\, 
\begin{pmatrix}
	|\mu|^2-\qt^2 \\
	-|\mu|^2
\end{pmatrix}
\bigg/\int d^2\qt\,\exp(-B\qt^2)\,(|\mu|^2+\qt^2) 
\,.\ee
Analytical values~: $D_{LL}=(\xi-1)/(\xi+1)$ and $D_{TT}= 
-\xi/(\xi+1)$ with $\xi=B|\mu|^2$. The geometrical decays of $|S_{n,z}|$ and $|\Sv_{n,T}|$ along the quark chain occur at different speeds. They are similar to the decays of charge and strangeness correlations. 
$D_{LL}$ and $D_{TT}$ saturate a Soffer-type \cite{inequal} positivity condition
\be \label{Soffer} 
 2|D_{TT}| \le 1+D_{LL}
\,.\ee
Indeed, $2D_{TT} = -1-D_{LL}$. This is due to the zero spin of $h_n$ (compare with text after Eq.(4.87) of \cite{inequal}). The negative value of $D_{TT}$ leads to Collins asymmetries of alternate signs, in accordance with the $^3P_0$ mechanism. It comes from the $\sigma_z$ vertex for pseudoscalar mesons. For \emph{scalar} mesons we replace $\sigma_z$ by $\un$. In this case $D_{TT}$ is positive, ~$q_{n-1}$ and $\bar q_{n}$ tend towards opposite 
sides and the Collins effect is small, except for $h_1$. This is also the prediction of the $^3P_0$ mechanism.

\section{Inclusion of spin-1 mesons}
\def\GVL{G_L}
\def\GVT{G_T}
\def\GAT{\tilde{G}_T}
For a $J^{PC}=1^{--}$ vector meson and the associated self-conjugate multiplet, the "minimal" emission vertex written with Pauli matrices is  
\be  \label{vectorVertex} 
\Gamma=\GVL\,V^*_z\, \un + \GVT\ \vsigma.{\vold{V}}^*_T\ \sigma_z 
\,,\ee
where $\Vv$ is the vector amplitude of the meson normalised to $\Vv.\Vv^*=1$. It is obtained from the relativistic 4-vector $V^\mu$ first by a longitudinal boost which brings the hadron at $p_z=0$, then a transverse boost which brings the hadron at rest. 

For a $J^{PC}=1^{++}$ axial meson of amplitude $\Av$, the "minimal" emission vertex is  
\be  \label{axialVertex} 
\Gamma= \GAT\,\vsigma.{\vold{A}}^*_T
\,.\ee
It differs by a $\sigma_z$ matrix from the second term of (\ref{vectorVertex}). A term of the form $\tilde{G}_L\,A^*_z\, \sigma_z$ with constant $\tilde{G}_L$ is not allowed by the forward-backward equivalence. 

Let us treat the case where the $1^{\rm st}$-rank particle is a $\rho^+$ meson and fix the momenta $p(\pi^+)$ and $p(\pi^0)$ of the decay pions. Then $V^\mu\propto p(\pi^+)- p(\pi^0)$, which is real, corresponding to a linear polarisation. Replacing the $\sigma_z$ coupling of (\ref{Ma}) by (\ref{vectorVertex}) we obtain 

\vspace{-5mm}
\begin{eqnarray} \label{rho} 
\cr I(\pv_T,\Vv) &=& \exp(-B\qt^2)\,|\GVT|^2 \times 
\cr & & \{\,(|\alpha|^2V_z^2+\Vv_T^2)\,(|\mu|^2+\qt^2)-4\Vv_T.\qt\,V_z\,\IM(\alpha)\IM(\mu) 
\cr &+& 2\IM(\mu)\,|\alpha|^2V_z^2\,\Sv.\qtil
\cr &+&  2 \IM(\alpha)\, (|\mu|^2+\qt^2) \,V_z\,\Vv_T.\tilde\Sv
\cr &+& 2 \IM(\mu)\,(\Vv_T.\qtil \,\Vv_T.\Sv  +{\Vv}_T.\qt \,{\Vv}_T.\tilde{\Sv})
\cr &+& 4\RE(\alpha)\IM(\mu)\,S_z \,V_z\, \Vv_T.\qtil \,\}~,
\end{eqnarray}
with $\qt\equiv\qt_1=-\pv_{T}(\rho^+)$ and $\alpha\equiv\GVL/\GVT$. Let us comment this formula~:

\ni$\bullet$~ The $2^{\rm nd}$ line is for unpolarised quark. It gives some tensor polarisation.

\ni$\bullet$~ The $3^{\rm rd}$ line is a Collins effect for the $\rho^+$ as a whole, opposite to the pion one (compare with (\ref{rank1}) and only for longitudinal linear polarisation, in accordance with the $^3P_0$ mechanism. For $\langle V_z^2\rangle=1/3$ (unpolarized $\rho^+$) and $\alpha=1$ one recovers the Czyzewski prediction \cite{Czyzewski} $A_T$(leading~$\rho$)/$A_T$(leading~$\pi$) $=-1/3$. 

\ni$\bullet$~ The $4^{\rm th}$ line gives an oblique polarisation in the plane perpendicular to $\Sv_T$ corresponding to $\hat h_{\bar 1}$ or $h_{1LT}$ of \cite{Ji,Bacchetta-M}. After $\rho^+$ decay, it becomes a relative $\pi^+-\pi^0$ Collins effect. 

\ni$\bullet$~ The $5^{\rm th}$ line is a new type of asymmetry, in $\sin[2\varphi(\Vv)-\varphi(\qt)-\varphi(\Sv)]$. 

\ni$\bullet$~ The last line also gives an oblique polarisation, but in the plane perpendicular to $\pv_T(\rho^+)$. After $\rho^+$ decay, it becomes jet-handedness. Indeed, ignoring an effect of transverse boost, we have $\Vv_T\propto\pv_T(\pi^+)-\pv_T(\pi^0)$, therefore $\Vv_T.\qtil\propto-\pv_T(\pi^+)\times\pv_T(\pi^0)$. 

\section{Conclusion}

For the direct fragmentation of a transversely polarised quark into pseudo-scalar mesons, 
the model we have presented has essentially one free complex parameter $\mu$ and reproduces the results of the semi-classical $^3P_0$ mechanism~: large asymmetry for the $2^{\rm nd}$-rank meson, Collins asymmetries of alternate sides for the subsequent mesons. In addition, it possesses a jet-handedness asymmetry, generated in two steps~: partial transformation of helicity into transversity, then Collins effect. 

We have also considered the inclusion of spin-1 mesons. When longitudinally polarised, a leading vector meson has a Collins asymmetry opposite to that of a pseudoscalar, as also expected from the $^3P_0$ mechanism. The decay pions of a $\rho$ meson exhibit a relative Collins effects as well as jet-handedness. These effects are associated to oblique linear polarisations of the $\rho$ meson. The fact that two pions coming from a $\rho$ show the same spin effects as two successive "direct" pions is reminiscent of duality. 

Even for unpolarised initial quarks, the spin degree of freedom of the cascading quark has to be considered. It enhances the $\langle\pv_T^2\rangle$ of the pseudoscalar mesons compared to scalar and longitudinal vector mesons. 

A next task for building a realistic Monte-Carlo generator with quark spin is to take into account the $(\qt_{n-1}-\qt_{n})^2$ term in (\ref{shell}).
One must also be aware that there exist other mechanisms of spin asymmetries in jets. For example the Collins effect can be generated by the interference between direct emission and the emission via a resonance \cite{Collins-L}.

\end{document}